\def\year{2020}\relax
\documentclass[letterpaper]{article} 
\usepackage{aaai20}  
\usepackage{times}  
\usepackage{helvet} 
\usepackage{courier}  
\usepackage[hyphens]{url}  
\usepackage{graphicx} 

\usepackage{mathtools}

\usepackage{graphicx}  
\usepackage[ND,SEQ]{prftree}
\usepackage{amsthm}

\newtheorem{theorem}{Theorem}
\newtheorem{definition}{Definition}
\newtheorem{lemma}{Lemma}
\newtheorem{corollary}{Corollary}
\newtheorem{proposition}{Proposition}
\newtheorem{example}{Example}

\title{Deciding the Loosely Guarded Fragment and Querying Its Horn Fragment Using Resolution}
\author{Sen Zheng, Renate A. Schmidt\\
Department of Computer Science, University of Manchester\\ 
Oxford Road, Manchester M13 9PL, UK
}

\begin{document}
\maketitle
\begin{abstract}
We consider the following query answering problem: Given a Boolean conjunctive query and a theory in the Horn loosely guarded fragment, the aim is to determine whether the query is entailed by the theory. In this paper, we present a resolution decision procedure for the loosely guarded fragment, and use such a procedure to answer Boolean conjunctive queries against the Horn loosely guarded fragment. The Horn loosely guarded fragment subsumes classes of rules that are prevalent in ontology-based query answering, such as Horn $\mathcal{ALCHOI}$ and guarded existential rules. Additionally, we identify star queries and cloud queries, which using our procedure, can be answered against the loosely guarded fragment.
\end{abstract}

\section{Introduction}
Our motivation of considering query answering problem stems from ontology-based data access (OBDA) systems~\cite{xiao2018ontology}, which have attracted much recent attention in the knowledge representation and database communities. In particular, since the Horn loosely guarded fragment subsumes mainstream rules in OBDA systems, such as Horn $\mathcal{ALCHOI}$ \cite{baader_horrocks_lutz_sattler_2017} and the guarded existential rules (a.k.a. guarded TGDs) \cite{CALI201257}, our interest is the development of a practical procedure for answering queries against the Horn loosely guarded fragment \cite{van1997dynamic}.  

To retrieve information from OBDA systems, the main querying mechanisms are (Boolean) conjunctive queries. Given a Boolean conjunctive query (BCQ) $q$, a set $\Sigma$ of rules and a database $\mathcal{D}$, checking whether $\Sigma \cup \mathcal{D} \models q$ is equivalent to checking whether $\Sigma \cup \mathcal{D} \cup \lnot q \models \bot$, so that the problem of answering BCQ can be reduced to deciding satisfiability. Such BCQ answering problems can be recast as query containment/entailment/evaluation problems in database research \cite{BAGET20111620}, constraint satisfaction problems and homomorphism mapping problems in general AI research \cite{Vardi:2000:CSD:335168.335209}. Although finding answers for queries is also an important problem, Boolean conjunctive query answering is widely studied \cite{BAGET20111620,barany2010querying,cali2013taming,CALI201257,glimm2008conjunctive,10.1007/978-3-642-39212-2_27}. In this paper, we particularly focus on an open problem, namely BCQ answering for the Horn loosely guarded fragment.



The complexity of BCQ answering for the guarded fragment is 2EXPTIME-complete \cite{barany2010querying}, and satisfiability checking for the clique-guarded negation fragment, which subsumes both BCQs and the loosely guarded fragment, is also 2EXPTIME-complete~\cite{Barany:2015:GN:2799630.2701414}. These complexity results show that BCQ answering for the Horn (loosely) guarded fragment is decidable, however, as yet there is no practical (i.e., implementable) procedure.


Let us give a quick review of our settings and explain why we start our investigation for query answering with deciding the loosely guarded fragment (LGF). A \emph{loosely guarded quantified formula} \cite{van1997dynamic,10.2307/2586808} has the form $\forall \overline x (G_1 \land \ldots \land G_n \rightarrow F)$ where $G_1, \ldots, G_n$ are atoms that are called \emph{guards}, $F$ is a loosely guarded formula where i) all free variables of $F$ occur in $G_1 \land \ldots \land G_n$, and ii) the variables in $G_1, \ldots, G_n$ are either free, or they co-occur with each other in a $G_i$ among the $G_1, \ldots, G_n$. The Horn fragment of LGF will be referred to Horn LGF. A \emph{Boolean conjunctive query} is a first-order formula of the form $q = \exists \overline x \varphi(\overline x)$ where $\varphi$ is a conjunction of atoms containing only constants and variables. One can obtain a \emph{query clause} $Q$ by simply negating a BCQ. Hence, $Q$ is a negative clause containing no compound terms.

To answer BCQs over Horn LGF, we begin with deciding~LGF, since it can be observed that i) some loosely guarded formulae are naturally cyclic BCQs \cite{Bernstein:1981:USS:322234.322238} because a loosely guarded formula $F$ allows multiple guards, and all variables in the clausal form of $F$ (loosely guarded clauses) co-occur with each other in one of the guards. E.g., the loosely guarded formula $F = \exists x y z (A_1xy \land A_2yz \land A_3xz)$ is a cyclic BCQ. In fact, it shows that the conjunctive queries with the hyper-tree width property are strongly connected to LGF \cite{GOTTLOB2003775}; ii) a loosely guarded formula allow variables chaining multiple literals ({\emph{chained variables}}), as in $F$, which can cause term depth increase during reasoning (see Example 1). We see that understanding the handling of chained variables in LGF helps us handle chained variables in query clauses. Hence, in this paper, we first provide a decision procedure to decide~LGF. Subsequently we show that such a procedure can be extended to answer BCQs over Horn LGF and answer restricted BCQs over LGF. 

Considering the only existing decision procedure for answering BCQs over (Horn) LGF essentially aims at the theoretical analysis \cite{Barany:2015:GN:2799630.2701414}, our focus is on devising a practical decision procedure, so that one can implement our procedure as query engines. We build our procedure using resolution in the framework of \cite{DBLP:books/el/RV01/BachmairG01}, which is a standard in the area of automated reasoning and provides basis for most first-order logic reasoners such as Spass \cite{weidenbach2009spass}, Vampire \cite{riazanov2001vampire} and E \cite{Schulz:LPAR-2013}. References for using resolution with refinement as practical decision procedures are \cite{ganzinger1998resolution,ganzinger1999superposition,hustadt1999maslov,hustadt1999resolution,hustadt1997evaluating,bachmair1993superposition}.
 
%

In resolution-based reasoning, the main challenges to avoid non-termination are: i) avoiding unlimited growth of the number of literals in resolvents, and ii) avoiding unlimited growth of term depth in resolvents. The former can be tackled by using the  property of loosely guarded clauses $C$: Guards in $C$ contain all variables of $C$, so that using our inference system, the number of literals cannot increase indefinitely in the resolvents. As for the latter, consider this example:

\begin{example} 
$Q$ is a query clause and $C_1, C_2$ are loosely guarded clauses:
\begin{align*}
&Q = \lnot A_1(x,y) \lor \lnot A_2(y,z), \\ &C_1 = A_1(f(x_1,y_1),x_1) \lor B_1(g(x_1,y_1)) \lor \lnot G_1(x_1,y_1), \\
&C_2 = A_2(h(x_2,y_2),x_2) \lor \lnot G_2(x_2,y_2)
\end{align*}
Performing resolution among $Q, C_1, C_2$ derives $R = B_1(g(h(x_2,y_2)),y_1) \lor \lnot G_1(h(x_2,y_2),y_1) \lor \lnot G_2(x_2,y_2) $. In $R$, $g(h(x_2,y_2))$ is deeper than all the terms in $Q, C_1$ and $C_2$. This happens when a query clause $Q$ contains a chained variable $y$ where: i) $y$ needs to be unified with a variable $x_1$ in $C_1$ and a non-ground compound term $h(x_2,y_2)$ in $C_2$ at the same time and, ii) $x_1$ occurs in a non-ground compound term $g(x_1,y_1)$ in $C_1$. However, such term depth increase can be avoided if we perform resolution on $Q$ and $C_1$ first. We introduce the top variable technique in its corresponding section to show how the term depth increase in $R$ can be prevented. 
\end{example}

Our resolution decision procedure for LGF is a variation of procedures presented in \cite{de2003deciding} and \cite{ganzinger1999superposition}. Like \cite{ganzinger1999superposition}, but unlike \cite{de2003deciding}, which uses a non-liftable ordering, our procedure uses admissible and liftable orderings with selection, and is consistent with the resolution framework of \cite{DBLP:books/el/RV01/BachmairG01}. Inspired by the `MAXVAR' technique in \cite{de2003deciding} and the partial hyper-resolution technique from \cite{ganzinger1999superposition}, we use the top variable technique to avoid term depth increase. \cite{ganzinger1999superposition} mainly focuses on deciding the guarded fragment, and refers to a manuscript version of \cite{de2003deciding} for technical details about using the `MAXVAR' technique to decide LGF. As for `MAXVAR', unlike \cite{de2003deciding}, we use a unification-first approach to identify top variables while \cite{de2003deciding} finds `MAXVAR' by variable depth first, then applies a specific unification algorithm. It turns out that our approach allows top variables~(`MAXVAR') being easily identified since no specific permutation and unification algorithms are needed. Further, we embed the top variable technique into the framework of~\cite{DBLP:books/el/RV01/BachmairG01} as a selection function~(top selection) with completeness proofs, so that, together by using liftable orderings, one immediately benefits from being able to use the notions of redundancy in that framework. Additionally, we generalise the pre-conditions of `MAXVAR', so that the top technique can be applied to a larger class than LGF (any clauses that satisfy conditions of query pair clauses), including queries, so that one can use the top technique to query other fragments of first-order logic to avoid variable depth growth as well. This makes our procedure applicable for the problems of BCQ answering for Horn LGF, and the star/cloud query answering for LGF.

Without hurting the result of this paper, we discuss variable depth, rather than term depth for the termination result. This holds because a term depth can grow infinitely if and only if a variable depth grows infinitely.

The contributions of this paper are:
\begin{itemize}
\item A variation of the resolution-based decision procedure for LGF in \cite{de2003deciding}, situated in the framework of \cite{DBLP:books/el/RV01/BachmairG01}.
\item By expanding the top variable technique to query pair clauses, this procedure provides the basis for a practical decision procedure for answering BCQs over Horn LGF.
\item We identify star queries and cloud queries so that one can use our procedure to answer these queries against LGF.
\end{itemize}
\section{Preliminaries}
Let \textbf{C}, \textbf{F}, \textbf{P} denote pairwise disjoint sets of \emph{constant symbols}, \emph{function symbols} and \emph{predicate symbols}, respectively. The definition of \emph{(compound/ground) term}, \emph{atom}, \emph{literal}, \emph{clause}, \emph{expression}, \emph{argument}, \emph{unifier}, \emph{most general unifier} (mgu) and \emph{simultaneous mgu} are defined as usual in automated reasoning (see e.g., \cite{DBLP:books/el/RV01/BachmairG01} for details). A literal $L$ is a \emph{non-ground compound literal} if~$L$ contains at least one non-ground compound term. Let $\overline x$, $\overline {A}$, $\mathcal{C}$ denote a sequence of variables, a sequence of atoms and a set of clauses, respectively. Let $var(t)$, $var(C)$ and $var(\overline {A_n})$ denote sets of variables in a term $t$, a clause $C$ and a sequence of atoms $\overline {A_n}$, respectively.


The variable depth of a term $t$, denoted as $vdp(t)$, is defined as follows: i) if $t$ is ground, then $vdp(t)=-1$, and if~$t$ is not ground, then ii) if $t$ is a variable, then $vdp(t) = 0$, and iii) if $t$ is a non-ground compound term $f(u_1, \ldots, u_n)$, then $vdp(t) = 1 + max(\{vdp(u_i) \ | \ 1 \leq i \leq n \})$. A term~$t$ is \emph{flat} if $vdp(t) \leq 0$. A term $t$ is \emph{simple} if $vdp(t) \leq 1$.
A \emph{flat} (\emph{simple}) \emph{atom, literal} and \emph{clause} is an atom, a literal and a clause such that every term in it is flat (simple). We say a term $t$ is a simple non-ground compound term if $vdp(t) = 1$.
Assume $C = \lnot A_1 \lor \ldots \lor \lnot A_n \lor D$ is a clause where $\lnot A_1 \lor \ldots \lor \lnot A_n$ are flat literal. Then $x \in var(A_1, \ldots, A_n)$ can be: i) a \emph{chained variable}: $x$ occurs in two literals $A_i, A_j$ among $A_1, \ldots, A_n$ such that $var(A_i) \not \subseteq var(A_j)$, $var(A_j) \not \subseteq var(A_i)$, and $x \in var(A_i) \cap var(A_j)$, and ii) an \emph{isolated variable}: $x$ is not chained. In $Q$ of Example 1, $y$ is a chained variable and $x,z$ are isolated variables. By the \emph{length} of a clause, we mean the number of literals that occur in a clause, and by the \emph{depth} of a clause, we mean the deepest variable depth of a clause. In this paper, we assume the input clauses (formulae) are of fixed-length and fixed-width.

A \emph{weakly covering} term is a compound term $t$ such that for every non-ground, compound subterm $s$ of $t$, it is the case that $var(s) = var(t)$ \cite{FLTZ93}. A literal $L$ is weakly covering if each argument of $L$ is either a ground term, a variable, or a weakly covering term $t$, such that $var(t) = var(L)$. A clause $C$ is weakly covering if each term $t$ in~$C$ is either a ground term, a variable, or a weakly covering term such that $var(t)=var(C)$. E.g., the clause $C_1 = \lnot A_1(fx y z a, x, y, ga) \lor A_2xyz$ is a weakly covering clause since the only non-ground compound term $fx y z a$ satisfies that $var(fx y z a) = var(C_1)$, however, the clause $C_2 = \lnot A_1(gy, y, ga) \lor A_2(hxy)$ is not weakly covering since $var(gy) \neq var(C_2)$. Here the notion of weakly covering literals in \cite{FLTZ93} is extended to weakly covering clauses; reasons are given in the loosely guarded clauses section.


Recall, the \emph{rule set} $\Sigma$ denotes a set of first-order formulae and the \emph{database} $\mathcal{D}$ denotes a set of ground atoms. A \emph{Boolean conjunctive query} is a first-order formula of the form $q =~\exists \overline x \varphi(\overline x)$ where $\varphi$ is a conjunction of atoms containing only constants and variables. We use the symbol $Q$ to denote the \emph{query clause} $\lnot q$, so that we can answer BCQ satisfiability of $\Sigma \cup D \models q$ by checking whether $\Sigma \cup D \cup Q \models \bot$.

\section{The Loosely Guarded Fragment}
\begin{definition}
\emph{The loosely guarded fragment (LGF)} is a fragment of first-order logic without equality and function symbols, defined inductively as follows: 
\begin{enumerate}
\item $\top$ and $\bot$ are in $LGF$.
\item If $A$ is an atom, then $A$ is in $LGF$.
\item $LGF$ is closed under Boolean combinations.
\item If $F \in LGF$ and $G_1$, $\ldots$, $G_n$ are atoms, then a formula $\forall \overline x (G_1 \land \ldots \land G_n \rightarrow F)$ belongs to $LGF$ if i)~all free variables of $F$ belong to $var(G_1, \ldots, G_n)$, and ii) for each variable $x \in \overline x$ and each variable $y \in var(G_1, \ldots, G_n)$ where $x \neq y$, $x$ and $y$ co-occur in a $G_i$. The negative literals $\lnot G_1, \ldots, \lnot G_n$ are called the \emph{guards} of this formula. 
\end{enumerate}   
\end{definition}
The first-order logic translation of a temporal logic formula $P$ \textbf{until} $Q$ is a loosely guarded formula: $\exists y (Rxy \land Qy \land \forall z((Rxz \land Rzy)\rightarrow Pz)))$, but the transitivity formula $\forall xyz ((Rxy \land Ryz) \rightarrow Rxz)$ is not a loosely guarded formula since $x$ and $z$ do not co-occur in a guard.

We use the loosely guarded formula $F$ in Example 2 to illustrate the clausal normal form transformation for LGF. 
\begin{example}
$\exists y (Rxy \land Qy \land \forall z((Rxz \land Rzy) \rightarrow \exists xPxy))$  
\end{example}


\section{The Resolution Calculus}
\label{calculus}

 %
 In this section, we introduce the resolution calculus, which gives us the main termination result of this paper. The inference steps are restricted by an admissible ordering and a selection function, so that the search space can be reduced when a reasoner computes inferences. For more technical details about rules used in this paper, we refer readers to~\cite{DBLP:books/el/RV01/BachmairG01}.

Let $\succ$ be a strict ordering, called a \emph{precedence}, on the symbols in the \textbf{C}, \textbf{F} and \textbf{P}. An ordering $\succ$ is \emph{liftable} if for all expressions $E_1$ and $E_2$ and all substitutions $\sigma$, $E_1 \succ E_2$ implies $E_1\sigma \succ E_2\sigma$. An ordering $\succ$ on literals is \emph{admissible}, if i) it is well-founded and total on ground literals, and \emph{liftable}, ii) $\lnot{A} \succ A$ for all ground atoms $A$, and iii) if $B \succ A$, then $B \succ \lnot A$ for all ground atoms $A$ and $B$. A literal $L$ is $\succ$\emph{-maximal with respect to a clause} $C$ if for any~$L^\prime$ in $C$, $L^\prime \not \succ L$, and $L$ is \emph{strictly} $\succ$\emph{-maximal with respect to a clause} $C$ if for any $L^\prime$ in $C$, $L^\prime \not \succeq L$. A \emph{selection function} $\mathcal{S}$ selects a possibly empty set of occurrences of negative literals in a clause $C$ with no restriction imposed. Inferences are only performed on eligible literals. A literal $L$ is \emph{eligible} in a clause $C$ if either nothing is selected in the selection function $\mathcal{S}$ and $L$ is a $\succ$-maximal literal with respect to $C$, or $L$ is selected by $\mathcal{S}$.

Inferences are computed using the following rules:

\textbf{Deduction:} $N$ derives $N, C$ if $C$ is either a resolvent or a factor of clauses in the set $N$.

Factors and resolvents are derived using:

\textbf{Ordered factoring:} $C \lor A_1 \lor A_2$ derives $(C \lor A_1)\sigma$, where i) $\sigma$ is the mgu of $A_1$ and $A_2$, and ii) no literal is selected in~$C$, and iii) $A_1\sigma$ is $\succ$-maximal with respect to~$C\sigma$.

\textbf{Ordered resolution with selection:} $\lnot A_1 \lor \ldots \lor \lnot A_n \lor D$, $B_1 \lor D_1, \ldots, B_n \lor D_n$ derive $(D_1 \lor \ldots \lor D_n \lor D) \sigma$ where i) either $\lnot A_1 \lor \ldots \lor \lnot A_n$ are selected in $D$, or $n=1$, no literal is selected, and $\lnot A_1\sigma$ is $\succ$-maximal with respect to $D\sigma$, and ii) no literal is selected in $D_1, \ldots, D_n$ and $B_1\sigma, \ldots, B_n\sigma$ are strictly $\succ$-maximal with respect to $D_1\sigma, \ldots, D_n\sigma$, respectively, and iii) $\sigma$ is a simultaneous mgu such that $A_1\sigma = B_1\sigma, \ldots, A_n\sigma = B_n\sigma$, and iv)~$\lnot A_1 \lor \ldots \lor \lnot A_n \lor D, B_1 \lor D_1, \ldots, B_n \lor D_n$ are pairwise variable-disjoint.

In ordered factoring and ordered resolution, maximality is computed using \emph{a-posteriori application} of the mgu $\sigma$. This means the maximal literal is determined after application of $\sigma$, derived from the unification algorithm applied to the premises of a rule. If the maximal literal is determined before the application of $\sigma$, we call this \emph{a-priori application}.

Redundancy is eliminated using:

\textbf{Deletion:} $N, C$ derives $N$ if $C$ is a tautology, or $N$ contains a variant of $C$, or $N$ contains a condensed form of $C$.

The `Deletion' rule is the only rule used to eliminate redundancy, and turns out to be sufficient for the termination result. Since we employ an admissible ordering with a selection function as resolution refinement in accordance with the framework of \cite{DBLP:books/el/RV01/BachmairG01}, we can also use more sophisticated simplification rules and redundancy elimination of that framework, e.g, subsumption deletion and forward/backward subsumption.

A ground clause $C$ is \emph{redundant with respect to} $N$ if there are ground instances $C_1\sigma, \ldots, C_n\sigma$ of clauses in $N$ such that $C_1\sigma, \ldots, C_n\sigma \models C$ and for each $i$, $C \succ C_i\sigma$. A non-ground clauses $C$ is \emph{redundant with respect to} $N$ if every ground instance of $C$ is redundant with respect to $N$. A set of clauses $N$ is \emph{saturated up to redundancy} (with respect to ordered resolution and selection) if any inference from non-redundant premises in $N$ is redundant in $N$ \cite{DBLP:books/el/RV01/BachmairG01}.
 
\section{The Decision Procedure}
Now we can discuss the resolution procedure for LGF.
\subsection{Clausal Normal Form Translation LGF-Trans}
The clausal normal form transformation we use is similar to the one in \cite{de2003deciding} and \cite{ganzinger1999superposition}, but i) free variables are assumed to be existentially quantified since the focus is on satisfiability checking, and ii) prenex normal form and outer Skolemisation~\cite{nonnengart2001computing} are used. Though outer Skolemisation may introduce Skolem functions of higher arities than inner/standard Skolemisation, outer Skolemisation turns out to be critical to guarantee that output clauses have the weakly covering property.

We use \emph{LGF-Trans} to denote the clausal normal form transformation below. Using $F$ in Example 2, one can obtain a set of loosely guarded clauses via the following steps:
\begin{align*}
\intertext{i) Add existential quantifiers to all free variables in $F$:}
&\exists xy (Rxy \land Qy \land \forall z((Rxz \land Rzy) \rightarrow \exists x Pxy)).
\intertext{ii) Rewrite $\rightarrow$ and $\leftrightarrow$ using conjunction, disjunction and negation, and transform $F$ into negation normal form, obtaining the formula $F_{nnf}$:}
&\exists x y (Rxy \land Qy \land \forall z(\lnot Rxz \lor \lnot Rzy \lor \exists x Pxy)).
\intertext{iii) Apply optimised structural transformation to $F_{nnf}$, that introduces fresh predicate symbols ($Q_1$) for universally quantified subformulae ($\forall z(\lnot Rxz \lor \lnot Rzy \lor \exists Pxy)$), obtaining the formula $F_{str}$:}
\begin{split}
&\exists x y (Rxy \land Qy \land Q_1xy) \ \land 
\end{split}\\
\begin{split}
\forall xy(\lnot Q_1xy \lor \forall z (\lnot Rxz \lor \lnot Rzy \lor \exists x Pxy)).
\end{split}
\intertext{iv) Find $\exists x y \forall uvw \exists x^\prime((Rxy \land Qy \land Q_1xy) \land (\lnot Q_1uv \lor \lnot Ruw \lor \lnot Rwv \lor Px^\prime v))$ as the prenex normal form of $F_{str}$, and apply outer Skolemisation: if $\forall \overline x$ is the subsequence of all universal quantifiers of the $\varphi$-prefix of subformula $\exists y \varphi$ of $\varphi$, then $\varphi[y/f(\overline x)]$ is the outer Skolemisation of $\exists y \varphi$. Skolem terms $a,b,fxyz$ are introduced, obtaining $F_{sko}$:}
\begin{split}
&Rab \ \land \ 
Qb \ \land \ 
Q_{1}ab \ \land 
\end{split}\\
\begin{split} 
\forall xyz(\lnot Q_1xy \lor \lnot Rxz \lor \lnot Rzy \lor P(fxyz,y))
\end{split}
\intertext{v) Drop all universal quantifiers and transform $F_{sko}$ into conjunctive normal form, obtaining loosely guarded clauses:}
\begin{split}
&Rab, \ Qb, \ Q_1ab, \ 
\lnot Q_1xy \lor \lnot Rxz \lor \lnot Rzy \lor P(fxyz, y)
\end{split}
\end{align*}

\subsection{Loosely Guarded Clauses}
We now describe loosely guarded clauses and their properties.
\begin{definition}
A \emph{loosely guarded clause (LGC)} $C$ is a clause satisfying the following conditions: 
\begin{enumerate} 
\item $C$ is simple and weakly covering, and
\item if $C$ is non-ground, then there is a set of negative literals $\lnot G_1, \ldots, \lnot G_n$ in $C$ that are flat. Then $\lnot G_1, \ldots, \lnot G_n$ are called the \emph{guards} of $C$, such that each pair of variables in $C$ co-occur in at least one of the guards.
\end{enumerate}
\end{definition}

We can immediately see that a ground clause is an LGC.
\begin{proposition}
Using \emph{LGF-Trans}, every loosely guarded formula can be transformed into a set of LGCs.
\end{proposition}
The class of LGCs strictly subsumes LGF since function symbols are allowed. If using an admissible ordering in which function symbols have higher precedence than other symbols, then non-ground compound terms in an LGC $C$ are always larger than variables in $C$ due to the weakly covering property. E.g., with a lexicographic path ordering $\succ_{lpo}$ \cite{dershowitz1982orderings}, considering an LGC $C = \lnot A_1xy \lor \lnot A_2yz \lor \lnot A_3xz \lor D(fxyz)$ and an arbitrary substitution $\sigma$, $D$ is $\succ_{lpo}$-maximal with respect to $C$ if $D\sigma$ is $\succ_{lpo}$-maximal with respect to $C\sigma$ since $var(D) = var(C)$. This shows that when determining the maximal literal in an LGC, the result of a-priori application follows the result of the a-posteriori application. To avoid the overhead of pre-computating the mgu using the a-posteriori application \cite{FLTZ93}, we use the a-priori application. We show this result in Lemma 1:
\begin{lemma}
Assume $C$ is a weakly covering clause containing a non-ground compound literal $L$ and $\sigma$ is an arbitrary substitution. Using any admissible ordering $\succ$ with a precedence that function symbols are larger than other symbols, $L$ is $\succ$-maximal with respect to $C$ if $L\sigma$ is $\succ$-maximal with respect to $C\sigma$.
\end{lemma}
An obvious property of a weakly covering clause $C$ is that all non-ground terms and literals in $C$ are also weakly covering. Formally stated as:
\begin{lemma}
If a clause $C$ is weakly covering, then for each non-ground compound term $t$ and each non-ground compound literal $L$ occur in $C$, $var(t) = var(L) = var(C)$.
\end{lemma}

\subsection{The Top Variable Technique}
Before discussing the resolution calculus for LGCs, we introduce the top variable technique, as a variation of the `MAXVAR' technique in \cite{de2003deciding}. The top variable technique is a look-ahead approach to prevent variable depth increase in the resolvents: Suppose we have a set of clauses that can lead to variable depth increase in the resolvent. Using the top variable technique, we first identify clauses that lead to the potentially deepest terms, and then perform resolution on those clauses first. Next we perform inference on the rest of the clauses. In such a manner of performing resolution, we show that no variable depth increase occurs in the resolvents.

\begin{definition} [Query Pair Clauses]
Let $\overline {A_n}$, $\overline {B_n}$ be a sequence of atoms $A_1, \ldots, A_n$ and a sequence of weakly covering atoms $B_1, \ldots, B_n$, respectively. $(\overline {A_n}$, $\overline {B_n})$ is a \emph{query pair} if they satisfy these conditions:
\begin{enumerate}
\item $\overline {A_n}$ is flat and non-ground, and $\overline {B_n}$ is simple.
\item Each $B_i \in \overline {B_n}$ either is a non-ground compound literal or is a ground literal.
\item $var(\overline {A_n}) \cap var(\overline {B_n}) = \emptyset$, and $B_1, \ldots, B_n$ are pairwise variable disjoint.
\item There exists an mgu (simultaneous mgu if $n > 1$) $\sigma$ such that for each $A_i \in \overline {A_n}$, $B_i \in \overline {B_n}$, $ A_i\sigma = B_i\sigma$.
\end{enumerate}
Let $(\overline {A_n}$, $\overline {B_n})$ be a query pair. \emph{Query pair clauses} for a query pair $(\overline {A_n}$, $\overline {B_n})$ is a set of clauses: $C= \lnot A_1 \lor \ldots \lor \lnot A_n \lor D$, $C_1 = B_1 \lor D_1$, $\ldots$, $C_n = B_n \lor D_n$ where $D$ is a flat clause and $D_1, \ldots, D_n$ are simple clauses. 
\end{definition}
From now on, we also use mgu to denote the simultaneous mgu. To find top variables in $\overline {A_n}$ in a query pair, one needs to find the mgu between $\overline {A_n}$ and $\overline {B_n}$ to identify the variable orderings over $var(\overline {A_n})$. 
\begin{definition} [Variable Ordering]
Let $(\overline {A_n}$, $\overline {B_n})$ be a query pair and let an mgu $\sigma$ satisfy Condition 4 in Definition 3. 

By $>_v, =_v$ we denote a \emph{variable ordering} over $var(\overline {A_n})$, which is defined by: for $x, y \in var(\overline {A_n})$, i) $x >_v y$ iff $vdp(x\sigma) > vdp(y\sigma)$, ii) $x =_v y$ iff $vdp(x\sigma) = vdp(y\sigma)$.
\end{definition}
Using the notion of variable orderings, we define top variables, and show the existence of top variables in query pairs:
\begin{definition} [Top variable]
\label{topVar}
Given a query pair $(\overline {A_n}$, $\overline {B_n})$, a variable $x \in var(\overline {A_n})$ is a \emph{top variable} iff for each $y \in var(\overline {A_n})$, $x >_v y$ or $x =_v y$.
\end{definition}

\begin{proposition}
Let $(\overline {A_n}$, $\overline {B_n})$ be a query pair. Then at least one of the variables in $\overline {A_n}$ is a top variable.
\end{proposition}
The idea behind the top variable technique is finding the potentially deepest term of query pair clauses. To realise it, we first apply the unification algorithm, then make the literals in the main premise containing the potentially deepest terms eligible literals. In Example 1, the mgu $\sigma = \{x/f(hx_2y_2, y_1),y/hx_2y_2,z/ x_2,x_1/ hx_2y_2\}$, thus applying resolution among $Q, C_1, C_2$ derives $B_1(g(hx_2y_2), y_1)$, in which the first argument is deeper than all terms in~$Q, C_1, C_2$. Now we use top variables to find the deepest terms. First we find top variables in $Q$: since $vdp(x\sigma) > vdp(y\sigma) > vdp(z\sigma)$, $x >_v y >_v z$. Since $x$ is the top variable (potentially the deepest term), we make the literal $A_1$ eligible since $x$ only occurs in $A_1$, then applying resolution on clauses $Q, C_1$, deriving the resolvent $C_3 = \lnot A_2(x_1,z) \lor B_1(gx_1y_1) \lor \lnot G_1(x_1,y_1)$. Though $C_3$ is not weakly covering, there is no variable depth increase in $C_3$. 


Let $(\overline {A_n}$, $\overline {B_n})$ be a query pair, and assume query pair clauses $C, C_1, \ldots, C_n$ such that $C = \lnot A_1 \lor \ldots \lor \lnot A_n \lor D$, as the \emph{main premise}, $C_1 = B_1 \lor D_1$, $\ldots$, $C_n = B_n \lor D_n$ as the \emph{side premises}, where $D$ is flat and $D_1, \ldots, D_n$ are simple. Assume $A_1, \ldots, A_t$ is a sequence of atoms containing top variables and the respective counterparts are $B_1, \ldots, B_t$, which occur in $C_1, \ldots, C_t$, respectively, and $\sigma$ is the mgu such that $A_i\sigma = B_i\sigma$ where $1 \leq i \leq t \leq n$. We denote \emph{Res} as an application of resolution among the clauses $C, C_1, \ldots, C_t$:
\begin{displaymath}
 \prftree[r]{$\scriptstyle$}
    {B_1 \lor D_1, \ldots, B_t \lor D_t}
    {\lnot A_1 \lor \ldots \lor \lnot A_{t} \lor \ldots \lor \lnot A_n \lor D}
    {(D_1 \lor \ldots \lor D_t \lor \lnot A_{t+1} \lor \ldots \lor \lnot A_{n} \lor D)\sigma}
\end{displaymath}
Given two expressions $A(\ldots, t, \ldots)$ and $B(\ldots, u, \ldots)$, we say $t$ \emph{matches} $u$ if the argument position of $t$ in $A$ is the same as the argument position of $u$ in $B$. We show how top variables in $var(\overline {A_n})$ match, the result is stated as:
\begin{lemma}
In an application of \emph{Res},
\begin{enumerate}
\item a top variable matches either a ground term or a non-ground compound term, and
\item a non-ground compound term matches a top variable.
\end{enumerate}
\end{lemma}
Based on the matching in Lemma 3, we now show properties of mgus in \emph{Res}:
\begin{lemma}
In an application of \emph{Res}, these conditions hold:
\begin{enumerate}
\item The mgu assigns to top variables either simple non-ground compound terms or ground terms.
\item The mgu assigns to non-top variables in the main premise either variables or ground terms. 
\item The mgu assigns to variables in the side premises either variables or ground terms.
\end{enumerate} 
\end{lemma}

Using Lemma 4, we give Theorem 1, which says that, for query pair clauses, only resolving literals that contain the potentially deepest terms does not lead to variable depth growth in the resolvents.
\begin{theorem}
In an application of \emph{Res}, no variable depth growth occurs in the resolvents of a set of query pair clauses. 
\end{theorem}
\subsection{Resolution Refinement LGC-Refine}
Now we formally describe the orderings and selection as refinement to decide LGF. One can use any admissible ordering that satisfies the conditions in \emph{LGC-Refine}. Here a lexicographic path ordering $\succ_{lpo}$ \cite{dershowitz1982orderings} is used.
\begin{definition} [LGC-Refine]
Let \emph{LGC-Refine} denote the refinement: A lexicographic path ordering $\succ_{lpo}$ based on a precedence $f>a>p$ for $f \in \emph{\textbf{F}}$, $a \in \emph{\textbf{C}}$ and $p \in \emph{\textbf{P}}$, and a selection function such that the following conditions hold:
\begin{enumerate} 
\item If a clause contains negative non-ground compound literals, then at least one of these literals is selected. 
\item If a clause contains no negative non-ground compound literal, but there are positive non-ground compound literals, then the maximality principle with respect to $\succ_{lpo}$ is applied to determine the eligible literals.
\item If a clause contains no non-ground compound literals, select all the negative literals containing top variables. 
\end{enumerate}
\end{definition}

We use \emph{top selection} to denote selection based on the top variable technique. Condition 3 in \emph{LGC-Refine} implies that top selection is imposed not only to guards. E.g., although~$\lnot B(x)$ is not a guard in $\lnot A(x,y) \lor \lnot B(x)$, top selection would select both $A$ and $B$ if $x$ is a top variable.

\subsection{The Resolution Calculus LGC-Res}
Now we discuss how resolution with \emph{LGC-Refine} performs over LGCs. We use the notation \emph{LGC-Res} to denote the calculus consisting of the following: the `Deduction' rule, ordered factoring and ordered resolution with selection refined by \emph{LGC-Refine}, and the `Deletion' rule. When applying \emph{LGC-Res}, the `Deletion' rule and the `Deduction' rule are used whenever they are applicable. As usual, we assume the input clauses (after condensation and modulo variable renaming) are a finite set of fixed LGCs.

First we discuss the ordered resolution rule. We use \emph{Res$^\prime$} to denote the resolution rule when one of the premises satisfy Condition 3 in \emph{LGC-Refine}.

Let $C = \lnot A_1 \lor \ldots\lor \lnot A_n \lor D$ be a flat LGC, as the \emph{main premise}, and $C_i = B_i \lor D_i$ be a set of LGCs, as the \emph{side premises}. Let $\overline {A_{t(n)}}, \overline {B_{t}}$ denote $A_1, \ldots, A_{t(n)}$ and $B_1, \ldots, B_{t}$ where $1 \leq t \leq n$, respectively. Using \emph{LGC-Refine}, \emph{Res$^\prime$} is performed as:
\begin{displaymath}
 \prftree[r]{$\scriptstyle$}
    {B_1 \lor D_1, \ldots, B_t \lor D_t}
    {\lnot A_1 \lor \ldots \lor \lnot A_{t} \lor \ldots \lor \lnot A_n \lor D}
    {(D_1 \lor \ldots \lor D_t \lor \lnot A_{t+1} \lor \ldots \lor \lnot A_{n} \lor D)\sigma}
\end{displaymath}
where i) $C$ is non-ground and $D$ is positive, ii) each $A_i \in~\overline {A_t}$ contains at least one top variable and each $B_i \in \overline {B_t}$ is strictly~$\succ_{lpo}$-maximal with respect to $C_i$, respectively, and  $\sigma$ is the mgu such that $A_i \sigma = B_i \sigma$ where $1 \leq i \leq t$, iii) $C$,~$C_1$, $\ldots$, $C_t$ are pairwise variable disjoint.

Since the premises in \emph{Res$^\prime$} (LGCs) satisfy conditions of query pair clauses, we can inherit results of \emph{Res}. The particularities in \emph{Res$^\prime$} are: i) all premises are weakly covering clauses rather than only literals are weakly covering, and ii) each premise contains a set of guards. 

First we show that using \emph{Res$^\prime$}, every resolvent is simple:
%
\begin{corollary}
In an application of \emph{Res$^\prime$}, the resolvents of a set of LGCs are simple clauses.
\end{corollary}
To show the resolvents in \emph{Res$^\prime$} are LGCs, we need to discuss some unique properties in \emph{Res$^\prime$} comparing to \emph{Res}:

\begin{lemma}
In an application of \emph{Res$^\prime$}, if we use notions from \emph{Res$^\prime$}, and let $x$ be a top variable in $A_1, \ldots, A_s$ ($s \leq t$). Then 
\begin{enumerate}
\item $var(A_1, \ldots, A_s) = var(C)$, and 
\item $var(x\sigma) = var(C\sigma)$, and 
\item $var(x\sigma) = var(y\sigma)$ if $x,y$ are distinct top variables.
\end{enumerate} 	
\end{lemma}
Now we can show the resolvents in \emph{Res$^\prime$} have are indeed LGCs: they are weakly covering and contain a set of guards.
\begin{lemma}
In an application of \emph{Res$^\prime$}, the resolvents of a set of LGCs are LGCs. 
\end{lemma}

It remains to consider other possibilities in \emph{LGC-Res}. In particular, we discuss situations that are not covered by \emph{Res$^\prime$} such that there is no premise satisfying Condition 3 in \emph{LGF-Refine}: the negative premise satisfies Condition 1 in \emph{LGF-Refine} or is ground. This is the case when the ordered resolution with selection is naturally reduced to a binary case. 

\begin{lemma}
In an application of \emph{LGC-Res}, the factors of LGCs are LGCs, and the resolvents of LGCs are LGCs.
\end{lemma}
Now we can show the main result of this section:
\begin{theorem}
Given a set of LGCs, using \emph{LGC-Res}, all inferred clauses are LGCs.
\end{theorem}

So far we have shown that using \emph{LGC-Res}, the resolvents of LGCs are LGCs. Since an LGC is a simple clause, there is no variable depth increase during the inference. We still need to consider that using \emph{LGC-Res}, the length of the resolvents cannot be infinitely long:
\begin{lemma} 
In an application of \emph{LGC-Res}, the number of variables in derived clauses is no more than the number of variables of one of the premises of these derived clauses.
\end{lemma}
\subsection{Refutational Completeness and Termination}
This section we give the refutational completeness and termination results of applications of \emph{LGC-Res} over LGCs. For refutational completeness result, we particularly show that the top selection used in \emph{LGC-Refine} is compatible within the framework of \cite{DBLP:books/el/RV01/BachmairG01}.
\begin{theorem} [Refutational Completeness]
Let $N$ be a set of clauses that are saturated up to redundancy under \emph{LGC-Res}, then $N$ is unsatisfiable iff $N$ contains the empty clause.
\end{theorem}

Let \emph{LGF-Res} denote the combination of the clausal transformation \emph{LGF-Trans} and the resolution calculus \emph{LGC-Res} with refinement \emph{LGC-Refine}. Now we give the first main result of this paper:
\begin{theorem}
\emph{LGF-Res} decides LGF.
\end{theorem}

\section{Querying Horn LGF and LGF}
In this section, we aim to check whether $\Sigma \cup \mathcal{D} \cup Q \models \bot$ where $\Sigma$ are formulae in (Horn) LGF, $\mathcal{D}$ is a set of ground atoms and $Q$ is a query clause. Because $\Sigma$ and $\mathcal{D}$ can be transformed into (Horn) LGCs using \emph{LGF-Trans}, the aim now is to check whether $\mathcal{C} \cup Q \models \bot$ where $\mathcal{C}$ is a set of (Horn) LGCs and $Q$ is a query clause. In particular, we assume $Q$ is a fixed query clause. We show that when either $Q$ is restricted to star/cloud queries, or $\Sigma$ is restricted to Horn LGF, our procedure guarantees termination.

Since a query clause $Q$ contains no non-ground compound terms, $Q$ satisfies Condition 3 in \emph{LGC-Refine}, thus no particular new refinement for $Q$ is needed. Hence we still can use \emph{LGC-Refine} as refinement for inference rules. However, \emph{LGC-Res} does not contain a rule to compute resolvents of a query clause and a set of LGCs. Using \emph{LGC-Refine}, a query clause $Q$ and a set of LGCs $C_1, \ldots, C_n$ satisfy conditions of query pair clauses, thus we apply \emph{Res} to compute resolvents of $Q$ and $C_1, \ldots, C_n$. We use \emph{Query-Res} to denote \emph{LGF-Res} and \emph{Res} using refinement \emph{LGC-Refine}. Since no positive factoring can be applied to query clauses, we only discuss how resolution is performed on query clauses. 

According to Theorem 1, the following result holds:
\begin{corollary}
In the application of \emph{Res}, there is no variable depth growth in the resolvents of a query clause and a set of LGCs, thus the resolvents are simple clauses.
\end{corollary}

\subsection{Querying Horn LGF}
It follows from Theorem 2 that using \emph{Query-Res}, the resolvents of a set of LGCs are LGCs. We now discuss the resolvent of a query clause and a set of LGCs. Corollary 2 shows that a resolvent $R$ of a query clause and a set of LGCs is a simple clause. However, this simple clause $R$ can be neither an LGC nor a query clause. In Example 1, one can obtain $C_3 = \lnot A_2(x_1,z) \lor B_1(gx_1y_1) \lor \lnot G_1(x_1,y_1)$ using \emph{Res} with \emph{LGC-Refine}. Although $C_3$ is simple, $C_3$ is neither an~LGC (not weakly covering), nor a query clause (not flat).

We observe that by disallowing multiple positive non-ground compound literals in LGCs, using \emph{Res}, the resolvent $R$ of a query clause and Horn LGCs is a negative clause containing no non-ground compound term, thus $R$ is a query clause. If we change $C_1$ in Example 1 to a Horn LGC $C_1^\prime = A_1(fx_1y_1,x_1) \lor \lnot G_1(x_1,y_1)$, then by applying \emph{Res} with refinement \emph{LGC-Refine}, since $x$ is a top variable, resolution between $Q, C_1^\prime$ derives $C_3^\prime =\lnot A_2(x_1,z) \lor \lnot G_1(x_1,y_1)$, which is a query clause. According to \emph{LGC-Refine}, further inference between $C_2$ and $C_3^\prime$ requires another positive premise that contains a positive literal $G_1$ that is either ground or contains non-ground compound terms. Hence, no inference is performed between $C_2$ and $C_3^\prime$.

We show that \emph{Query-Res} can decide $\mathcal{C} \cup Q \models \bot$ where $\mathcal{C}$ are formulae in Horn LGF and $Q$ is a query clause. Notice that ground atoms $\mathcal{D}$ are immediately in Horn LGF. A \emph{Horn loosely guarded clause} (Horn LGC) is an LGC that contains at most one positive literal. The \emph{Horn loosely guarded fragment} (Horn LGF) is a subset of LGF that can be transformed into a set of Horn LGCs using \emph{LGF-Trans}.


Using \emph{Res}, when query pair clauses are query clauses and a set of Horn LGCs, the resolvents are query clauses: 


\begin{lemma}
In an application of \emph{Res}, the resolvents of Horn LGCs and a query clause are query clauses.
\end{lemma}

%
It turns out the proof of Lemma 9 does not require the premises to be guarded. Thus we can generalise Lemma 9 to a result such that the resolvents of a query clause and a set of simple, weakly covering Horn clauses are query clauses.

Since a Horn LGC contains at most one positive literal, ordered factoring cannot be applied. Now we show other possibilities of applying resolution in \emph{Query-Res}:
\begin{lemma}
Using \emph{Query-Res}, the resolvents of Horn LGCs are Horn LGCs.  
\end{lemma}
So far we showed that query clauses and Horn LGCs are closed under the inference system \emph{Query-Res}, and the variable depth of derived clauses does not increase. Now we consider the length of derived clauses:
\begin{lemma}
In an application of \emph{Query-Res}, given a finite set of fixed Horn LGCs and fixed query clauses, all derived clauses are fixed query clauses.    
\end{lemma}
Now we give the second main result of this paper:
\begin{theorem}
\emph{Query-Res} decides the problem of the BCQ answering for Horn LGF.
\end{theorem}

\subsection{Restricted Queries for LGF}
In this section, we answer loosely guarded queries, star queries and cloud queries over LGF. Theorem 4 implies that if a query clause $Q$ is expressible in LGCs, then using \emph{LGF-Res}, one can immediately answer $Q$ over LGF. E.g., one can answer a loosely guarded query $\exists xyz (Postgrad(x) \land citedBy(x,y)\land citedBy(y,z) \land citedBy(z,x))$ over LGF using \emph{LGF-Res}. This result is formally stated as:
\begin{corollary}
\emph{LGF-Res} decides the loosely guarded query answering problem for LGF.
\end{corollary}

%
Another observation from Example 1 is: The top variable $x$ does not occur with all other variables in $Q$. If $y$, which occurs with all other variables in $Q$, is a top variable, then using \emph{Res}, the resolvent among $Q, C_1, C_2$ is an~LGC. E.g., if we change $C_1$ to $C_1^\prime = A_1(x_1,fx_1y_1) \lor B_1(gx_1y_1) \lor \lnot G_1(x_1,y_1)$, to make $y$ a top variable, then using \emph{Res}, the resolvent of $Q, C_1^\prime, C_2$ is an LGC $B_1(gx_1y_1) \lor \lnot G_1(x_1,y_1) \lor \lnot G_2(x_1,y_1)$. This observation motivates our definition of star queries and cloud queries, which both guarantee the co-occurrence property between top variables and all other variables in the query clause. 

Before discussing star/could queries, we first give the definition of partners of a variable. The \emph{partner of a variable $x$} in a query $Q$ $par(x, Q)$ is a set of variables that co-occur with $x$ in an atom of $Q$, and $par(\overline x, Q)$ is interpreted as $par(x_1, Q) \cup \ldots \cup par(x_n, Q)$ where $x_i$ is in an atom of $Q$. E.g., Let a query $Q = \exists xyz (A_1xy \land A_2yz)$. Then $par(x, Q) = \{y\}$ and $par(x,z, Q) = par(x, Q) \cup par(z, Q) = \{y\}$. 

\begin{definition} [Star Query]
A BCQ is a \emph{star query} $Q$ if $Q$ contains a top variable $x$ such that $par(x, Q) = var(Q)$.   
\end{definition}
The notion of star query strictly extends that of loosely guarded query, since only one top variable need to occur with all other variables. E.g., a query clause $Q = \exists xyz (A_1xy \land A_2yz)$ is not a loosely guarded query since $x$ and $z$ do not co-occur in any literal in $Q$, but if $y$ is a top variable, $Q$ is a star query. 

\begin{definition} [Cloud Query]
A BCQ $Q$ is a \emph{cloud query} if $Q$ satisfies these conditions: 
\begin{enumerate}
\item $Q$ contains chained variables $\mathcal{V}$ that are top variables.
\item Each pair in $\mathcal{V}$ co-occur in an atom of $Q$.
\item $par(\mathcal{V}, Q) = var(Q)$.
\end{enumerate}    
\end{definition}    

The notion of cloud query is a further extension of that of a star query since a top variable does not need to occur with all other variables. An example of a cloud, but not star query is $Q = \exists xyzuv (A_1xy \land A_2yz \land A_3zuv \land A_4v)$ if $y,z$ are top variables. $Q$ is a cloud query because that $\mathcal{V} = \{y,z\}$, $y, z$ co-occur in $A_2$, and $par(\mathcal{V}, Q) = var(Q)$.

Unlike loosely guarded queries, star queries and cloud queries vary depending on whether the top variables co-occur with all other variables. 

Now we show that the resolvents of a set of LGCs and a star/cloud query are LGCs:
\begin{lemma}
Using \emph{Res} with refinement \emph{LGC-Refine}, the resolvents of a set of LGCs and a star/cloud query are LGCs.
\end{lemma}
%

Following the same idea of Lemma 8, we can show that the resolvents of a set of LGCs and a star/cloud query cannot have more variables than one of their side premises. Now we can state the third result of this paper:
\begin{theorem}
\emph{Query-Res} decides the problem of the loosely guarded query and star/cloud query answering for LGF.
\end{theorem}
\section{Conclusion and Future Work}
In this paper, we have presented, as far as we know, the first practical decision procedures for answering BCQs over Horn LGF. Inspired by the `MAXVAR' notion from \cite{de2003deciding}, we used the top variable technique to handle chained variables in query clauses. Based on this top variable technique, we showed the method \emph{LGF-Res} decides LGF, the method \emph{Query-Res} answers BCQs over Horn LGF, and it answers loosely guarded queries, star queries and cloud queries over LGF. This shows that \emph{Query-Res} provides essentials for the implementation of query answering over guard-related fragments as an extension for  existing first-order logic reasoners.

Using \emph{Query-Res}, an issue of answering BCQs against the whole of LGF is that: If the top variable is an isolated variable in a query clause $Q$, then the resolvents of $Q$ and a set of LGCs is neither a query clause nor an LGC (Example 1). Our next step is extending \emph{Query-Res} so that we can use it to answer BCQs against LGF, for which as yet there is no practical procedure. 


\section{Acknowledgements}
We very thank the anonymous reviewers for their useful comments.

\bibliographystyle{aaai}
\bibliography{aaai}
\end{document}